\documentclass[10pt]{article}

\usepackage{booktabs} % For formal tables

\newcommand{\stijn}[1]{}
\newcommand{\iman}[1]{}
\newcommand{\note}[1]{}

\usepackage{fullpage}
\date{}

\title{Incremental Techniques for Large-Scale Dynamic Query Processing}

\author{
Iman Elghandour$^1$, Ahmet Kara$^2$, Dan Olteanu$^2$, Stijn Vansummeren$^1$ \\ \\
$^1$Universit\'e Libre de Bruxelles \enspace\enspace $^2$University of Oxford \enspace\enspace
}

\begin{document}
\maketitle
\begin{abstract}
\noindent
Many applications from various disciplines are now required to analyze
fast evolving big data in real time. Various approaches for
incremental processing of queries have been proposed over the
years. Traditional approaches rely on updating the results of a
query when updates are streamed rather than re-computing these queries, and therefore, higher execution performance is expected.  However, they do not
perform well for large databases that are updated at high frequencies.  Therefore, new algorithms and approaches have been proposed in the literature to address these challenges by, for instance, reducing the complexity of processing updates. Moreover, many of these algorithms are now leveraging distributed streaming platforms such as Spark Streaming and Flink. In this tutorial, we briefly discuss legacy approaches for incremental query
processing, and then give an overview of the new challenges introduced due to
processing big data streams. We then discuss in detail the recently proposed algorithms that address some of these challenges. We emphasize the characteristics and algorithmic analysis of various proposed approaches and conclude by discussing future research directions.  
\end{abstract} 

\section{Introduction}
%\input{introduction}
%introduction about dynamic processing and a motivating example 
In a broad range of domains, such as Real Time Business Intelligence
and Complex Event Processing, contemporary applications require the
timely \emph{dynamic} processing of complex analytical queries on
continuously arriving data. Here, \emph{dynamic processing} refers to updating the query result, preferably in real-time, when the underlying data is updated.
Implementing such applications remains a difficult task, and involves
resolving two orthogonal challenges: 
\begin{itemize}
\item Designing a suitable dynamic query
processing algorithm that determines how the
application's query results are to be updated upon data changes,
taking into account that previous results are already available and
re-computation should be avoided to ensure timeliness. 
\item Designing an implementation and deployment of the selected dynamic
query processing algorithm that accounts for desiderata such as high
throughput, low latency, and the ability to process large data
sets. Current approaches mostly rely on distributed computing frameworks
such as MapReduce, Flink, Spark, or Storm, to achieve this.
\end{itemize}

Fortunately, in recent years, there has been a flurry of research on both
challenges that provide novel insights in how to resolve them. We
briefly survey these next.

\smallskip\noindent{\bf Algorithmical insights.}  Avoiding the
re-computation whenever an update is received has long been
approached using Incremental View Maintenance (IVM)
techniques~\cite{Chirkova:Views:2012:FTD,Gupta:1999:MVT:310709}. IVM
materializes the output of a query and then maintains that output
under updates.  Unfortunately, traditional IVM is not efficient for
large databases that are updated at a high frequency. Therefore, new
approaches have recently been proposed, whose objective is to reduce
the complexity of processing updates and/or to reduce the required
memory footprint.

%a brief overview of solutions from the literature that are centralized
Specifically, research in dynamic query processing has recently
received a big boost with: (1) the introduction of Higher-Order IVM
(HIVM)~\cite{DBT:VLDBJ:2014,DBLP:conf/sigmod/NikolicEK14,
  DBLP:conf/pods/Koch10}; (2) the identification of lower bounds and
worst-case optimal algorithms for processing
updates~\cite{berkholz-pods,berkholz-icdt,dan-icdt19}; (3) the practical
formulations of worst-case-optimal IVM that implement and extend these
algorithms~\cite{dyn-sigmod,dyn-vldb,factorized-ivm-sigmod}; and (4) the
introduction of the notion of differential dataflow for computations that require
recursive or iterative
processing~\cite{DistributedDataflow_commACM2016, DBLP:conf/cidr/McSherryMII13}. These approaches
often rely on materializing a succinct representation of a query's
output to maintain it more efficiently, and therefore present a
fundamental breakthrough with traditional IVM techniques.

\smallskip\noindent{\bf Big Data Frameworks support for dynamic query
  processing.}  Big data frameworks such as
MapReduce~\cite{MapReduce_OSDI2004} and Spark~\cite{SparkRDD_NSDI2012} are inherently
batch-oriented. Early approaches for implementing dynamic query
processing in these frameworks has focused on incremental processing
of MapReduce
tasks~\cite{Incoop_SoCC11,IncMR_CLOUD12,Nova_SIGMOD11}. These are,
however, based on traditional IVM techniques and suffer from high
latency of MapReduce and its open source implementation Hadoop. More
recent versions of distributed compute frameworks such as Apache
Spark~\cite{SparkStream_SOSP2013}, Apache
Flink~\cite{Stratosphere_VLDBJ2014}, and Twitter
Storm~\cite{STORM_SIGMOD2014} / Heron~\cite{Kulkarni:2015:THS:2723372.2742788} allow stream-based computations
instead of batch-based computations. Out of the box, these frameworks
mostly provide primitives for avoiding re-computation over sliding
windows, based on traditional IVM. In addition, they present low-level
programming primitives by which developers can express their own
dynamic query processing algorithms. More recently, there are
proposals to automatically incrementalize queries on distributed big
data frameworks.
Examples of these approaches include the distributed implementation of
HIVM~\cite{hivm-hotdog} and differential
dataflow~\cite{DistributedDataflow_commACM2016}, as well as Spark
Sructured Streaming~\cite{SparkStreaming_SIGMOD2018}.

%\section{Aims and scope}
%\input{aims_scope}
%
\section{Tutorial Structure}
%\input{structure}
%The tutorial will start by presenting the requirements of dynamic
%processing techniques as they emerge from real-world Big Data
%applications and the main challenges that need to be addressed. Then,
%we will briefly survey and summarize the traditional approaches to
%incremental query processing. Subsequently, we will discuss in detail
%the novel approaches referenced above, their algorithmic insights, and
%their characteristics.  Next, we discuss distributed approaches and
%frameworks. We finish by discussing open research problems.

The tutorial runs for 3 hours and is divided into the following four parts:

\smallskip \noindent
{\bf Part I: Introduction, desiderata, and traditional IVM}\\
We start the tutorial by giving an introduction to dynamic query
processing and show examples that motivate the need for efficient
incremental query processing. We give a high-level historical overview
of traditional approaches (known as First Order IVM) that have been
employed by conventional database systems to maintain query outputs.  We present the strong and weak points of traditional approaches and then discuss new challenges introduced by streaming large data at high frequencies.

\smallskip\noindent
{\bf Part II: Recent Algorithmic Advances in Dynamic
  Query Processing}\\
In the second part of the tutorial, we survey new
efficient approaches and algorithms for dynamic query processing.  We discuss the following research works: (1) Higher-Order IVM~\cite{DBT:VLDBJ:2014,DBLP:conf/sigmod/NikolicEK14,DBLP:conf/pods/Koch10}; (2) Complexity lower bounds for dynamic query
  processing~\cite{berkholz-pods,berkholz-icdt, dan-icdt19}; (3) Dynamic Yannakakis~\cite{dyn-sigmod,dyn-vldb}; (4) Factorized IVM~\cite{factorized-ivm-sigmod}; (5) Space-time tradeoffs~\cite{dan-icdt19}; and (6) Beyond conjunctive queries: relations over application-dependent rings~\cite{Green:2007:PS:1265530.1265535,factorized-ivm-sigmod, DBLP:conf/pods/Koch10}. 
  
\begin{sloppypar}
\smallskip\noindent
{\bf Part III: Dynamic Query Processing in
  Big Data Frameworks}\\
Incremental processing of
queries has been studied for queries executed by MapReduce~\cite{Incoop_SoCC11,IncMR_CLOUD12,Nova_SIGMOD11} and by other distributed
streaming platforms such as Spark Streaming~\cite{SparkStream_SOSP2013, SparkStreaming_SIGMOD2018}, Flink~\cite{Stratosphere_VLDBJ2014}, and Storm~\cite{STORM_SIGMOD2014}/Heron~\cite{Kulkarni:2015:THS:2723372.2742788}. However,
these systems rely on their users to specify how the queries are
maintained or employ traditional incremental view maintenance approaches.  Additionally, new parallel approaches that
are executed in distributed environments~\cite{hivm-hotdog} or that
extend incremental processing~\cite{DistributedDataflow_commACM2016} are
introduced. We discuss all the mentioned approaches and platforms
while highlighting the contributions that each one of them has made.
\end{sloppypar}

\smallskip\noindent
{\bf Part IV: Outlook}\\
Finally we conclude by summarizing the existing research solutions and
highlighting the open problems that are yet to be studied.

\section{Acknowledgment}
%Thanks to the  Wiener-Anspach foundation for the gracious support. 
%\noindent
%{\bf ACKNOWLEDGMENT}\\
The authors are graciously supported by the Wiener-Anspach foundation. The work has received funding from the European Union's Horizon 2020 research and innovation programme under grant agreement 682588.

%\bibliographystyle{abbrv}
%\bibliography{bibliography}

 \end{document}